\documentclass[runningheads]{llncs}

\usepackage{amsmath}
\usepackage{amssymb}
\usepackage[colorlinks,citecolor=blue,linkcolor=blue,urlcolor=blue]{hyperref}
\usepackage{enumitem}
\usepackage{multirow}
\usepackage{booktabs}
\usepackage{verbatim}
\usepackage{marvosym}

\usepackage[T1]{fontenc}
\usepackage{graphicx}

\begin{document}
\begin{sloppypar}

\title{Don’t Click the Bait: Title Debiasing News Recommendation via Cross-Field Contrastive Learning}
\titlerunning{TDNR-C²: Title Debiasing News Recommendation}

\author{Yijie Shu \and
Xiaokun Zhang \and
Youlin Wu \and
Bo Xu \and
Liang Yang \and
Hongfei Lin\textsuperscript{\Letter}}
\authorrunning{Y. Shu et al.}
\institute{School of Computer Science and Technology, Dalian University of Technology, Dalian 116024, China \\
\email{\{644581949,wuyoulin\}@mail.dlut.edu.cn} \\
\email{\{xubo,liang,hflin\}@dlut.edu.cn, dawnkun1993@gmail.com}}

\maketitle              % typeset the header of the contribution

\begin{abstract}
News recommendation emerges as a primary means for users to access content of interest from the vast amount of news. The title clickbait extensively exists in news domain and increases the difficulty for news recommendation to offer satisfactory services for users. Fortunately, we find that news abstract, as a critical field of news, aligns cohesively with the news authenticity. To this end, we propose a \textbf{T}itle \textbf{D}ebiasing \textbf{N}ews \textbf{R}ecommendation with \textbf{C}ross-field \textbf{C}ontrastive learning (TDNR-C²) to overcome the title bias by incorporating news abstract. Specifically, a multi-field knowledge extraction module is devised to extract multi-view knowledge about news from various fields. Afterwards, we present a cross-field contrastive learning module to conduct bias removal via  contrasting learned knowledge from title and abstract fileds. Experimental results on a real-world dataset demonstrate the superiority of the proposed TDNR-C² over existing state-of-the-art methods. Further analysis also indicates the significance of news abstract for title debiasing.

\end{abstract}

% 第一章：Introduction
\section{Introduction}
With the rapid growth of online news, individuals frequently encounter the dilemma of information overload while perusing news on websites such as Yahoo News and Google News \cite{NR_overload}. To solve this issue, news recommendation systems are designed to promptly select articles aligning with users' interest from a large amount of news data. With tremendous practical significance, news recommendations have garnered increasing attention recently \cite{NR_Intro1,NR_Intro2}.  

\begin{figure}[t]
\centering
\includegraphics[width=0.8\textwidth]{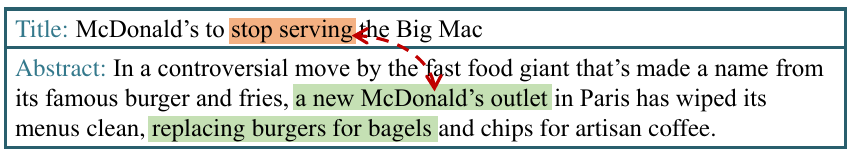} 
\caption{An example of news clickbait.} \label{fig1} 
\end{figure}

Unfortunately, extensive clickbait significantly impedes news recommendation effectiveness. Editors often use exaggerated titles to attract clicks, creating clickbait. As shown in Fig.~\ref{fig1}, a news article title on the ``Delicious''\footnote{https://www.delicious.com.au} website implies the cessation of McDonald's Big Mac sales, whereas a closer examination of the news abstract reveals that the article pertains to a coffee experiment conducted by a single McDonald's branch in France, rather than the global chain. This misleading title serves as a classic example of clickbait, initially drawing interest but ultimately eroding trust in news authenticity.

However, prevailing news recommendation methods often rely heavily on news titles for modeling, whether utilizing CNN \cite{DAN} or multi-head self-attention \cite{NPA,NRMS}. These approaches derive news representations solely from news titles, overlooking the potential clickbait issue within them, thus limiting the recommender's effectiveness.Fortunately, we have discovered that news abstracts better reflect the actual news content and offer richer information. As illustrated in Fig.~\ref{fig1}, news abstracts are crucial for providing authentic information when news titles exhibit clickbait tendencies. While some methods have attempted to integrate additional field information like abstracts instead of soly relying on titles \cite{MTRec,NAML,wu2020user}, they face challenges in effectively leveraging multi-field knowledge to combat clickbait. These challenges include the lack of field-specific knowledge extraction prior to aggregation, the introduction of noise at an early stage hindering clickbait elimination, and the limitations of simple attention or concatenation methods in identifying latent noise in news title, thereby inadequately addressing clickbait issues.

To address these challenges, we introduce the \textbf{T}itle \textbf{D}ebiasing \textbf{N}ews \textbf{R}ecommendation with \textbf{C}ross-field \textbf{C}ontrastive Learning (TDNR-C\textsuperscript{2}) framework. Leveraging contrastive learning as a self-supervised technique, we aim to reduce the noise in the original samples by comparing positive and negative samples. Considering that the alignment of abstracts with news facts, we compare these two fields to achieve the purpose of removing the noise of clickbait in news titles. Specifically, before aggregating multi-field information of news to obtain the news representation, we first learn the multi-view knowledge of each news field through a \textbf{M}ulti-\textbf{F}ield \textbf{K}nowledge \textbf{E}xtraction module (MFKE). Given the clickbait issue prevalent in the realm of title field, acquiring a comprehensive understanding of multi-view knowledge pertaining to news enables us to prevent the spreading of noise across all news fields. Then in the \textbf{C}ross-field \textbf{C}ontrastive learning module (C\textsuperscript{2}), we set the learned title-view knowledge and abstract-view knowledge of the same sample as positive samples, and between different samples as negative samples. This contrast allows us to capture semantic information and address the clickbait issue in news titles. At the same time, acknowledging the significant text length disparity between news titles and abstracts, we preemptively generated titles from news abstracts. Leveraging the T5 model\footnote{https://huggingface.co/Michau/t5-base-en-generate-headline} for its rich external knowledge, we aimed to bridge this gap and enhance data quality for contrastive learning.
% Experimental results confirm the method's performance enhancement and a 60\% reduction in abstract length, boosting the model's efficiency in utilizing abstract field knowledge.

In summary, the main contributions of our work are as follows:
\begin{itemize}[label=\textbullet]
\item We observe that the abstract is closer to the authenticity of news. It can effectively express the semantic information of the news, thereby contributing to the alleviation of clickbait issues.
\item In the TDNR-C² framework, we introduce two novel modules: MFKE and C². These modules effectively address the clickbait issue by contrasting different views of news knowledge extracted by MFKE.
\item Extensive experiments demonstrate the effectiveness of the proposed TDNR-C² for news recommendation. Further analysis also proves the significance of abstract in mitigating clickbait issue.
\end{itemize}

% 第二章相关工作
\section{Related Work}
\subsubsection{News Modeling in News Recommenders.} Traditional news recommendation methods, such as manual feature extraction models like Wide\&deep \cite{wide&deep} and SVM-based classifier \cite{SVM}, often demand substantial domain expertise and effort due to the manual crafting of features. These features may not adequately capture the semantic information present in news texts. With the advancements in natural language processing techniques, modern approaches leverage deep neural networks, like denoising autoencoder \cite{related_work2} and multi-head self-attention mechanism \cite{NRMS,NAML}, to model semantic content effectively. Despite these advancements, the premature integration of information from multiple fields can introduce noise, thereby limiting the news recommender's performance. Additionally, the lack of effective means to deal with the clickbait issue in news titles further restricts the performance of news recommenders. In the realm of session-based recommendation, several researches have addressed the challenge of integrating different types of knowledge through mechanisms such as bi-preference learning and co-guided learning \cite{PDM,DIMO,BiPNet}. To overcome the aforementioned challenges, we introduce a multi-field knowledge extraction module to prevent noise spread and a cross-field contrastive learning module to address clickbait issues in news titles. 
\subsubsection{Contrastive Learning for News Recommendation.} Contrastive learning, as a form of self-supervised learning strategy \cite{RW_CL1,RW_CL2}, enhances the model's representational capacity by comparing positive and negative samples between different views without additional labeled data. The crux lies in establishing positive and negative sample pairs, with most existing methods in the NLP domain employing data augmentation techniques to construct positive sample pairs \cite{RW_DataAug2,RW_DataAug3,RW_DataAug1}, such as masking, translation, or synonym replacement, while negative samples consist of samples from distinct categories or the remaining batch. These data augmentation methods do not fundamentally alter the semantic information of the text and struggle to effectively discern relevant information exaggerating news content in clickbait issue. Hence, this paper proposes a method of regenerating titles through news abstracts to address the aforementioned problem.
% 第三章：方法
\section{Methodology}
In this section, we first introduce the construction of interaction data and the necessary symbols. Then we show the TDNR-C² framework and discuss each module of TDNR-C² in detail. 
\begin{figure}
\includegraphics[width=\textwidth]{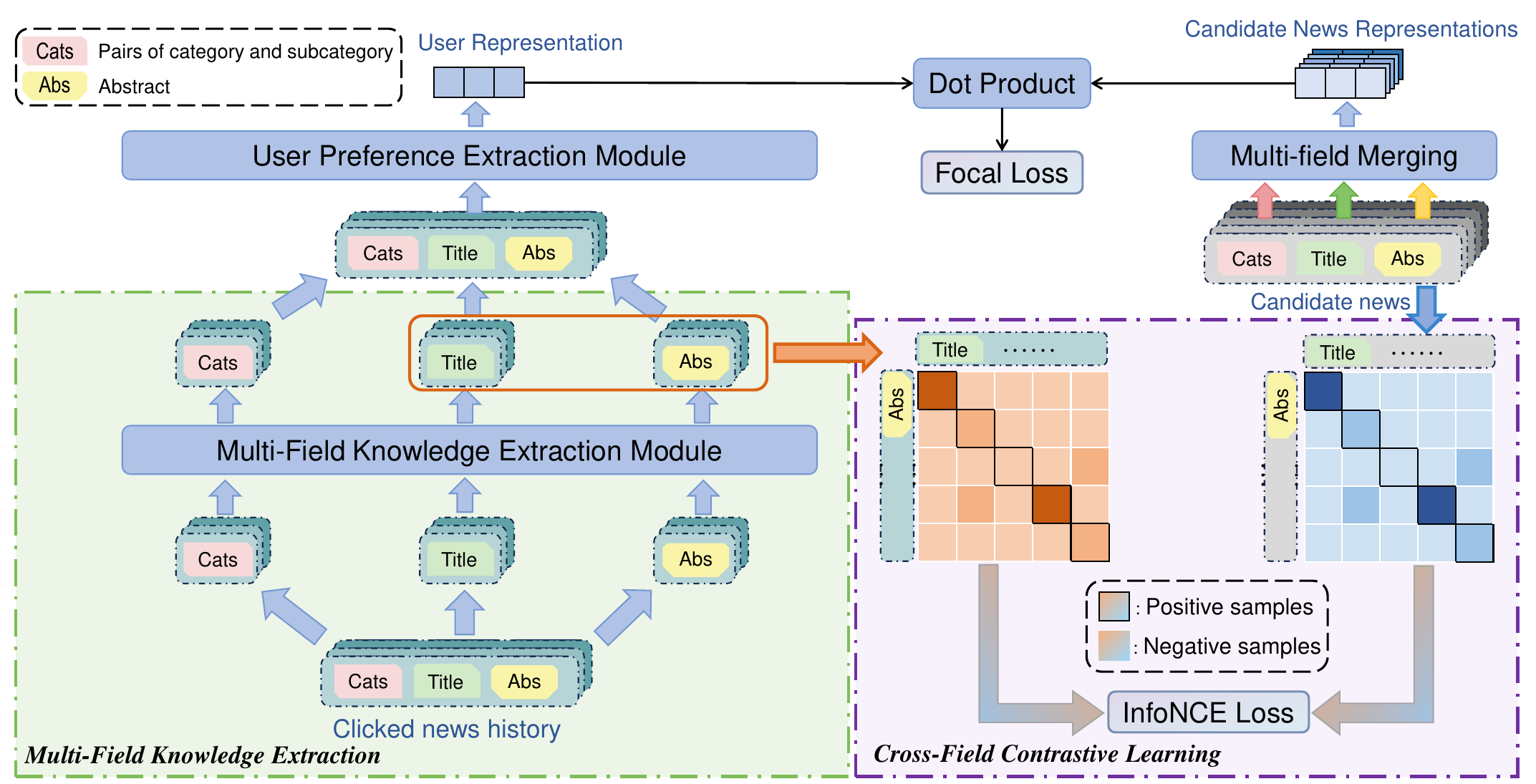}
\caption{The proposed TDNR-C² model comprises two key components: (1) MFKE, which converts clicked news history into field-level history and captures multi-view knowledge of the clicked news history; (2) C², a cross-field contrastive learning module designed for clicked news history and candidate news to capture semantic relevance between title and abstract, alleviating the clickbait issue in news titles.} \label{fig2}
\end{figure}
\subsection{Notations and Definitions}
Given \(U = \left\{ u_{1},\ldots,u_{v},\ldots,u_{V}, \right\}\) as the set of user behavior logs and \(N = \left\{ n_{1},\ldots,n_{w},\ldots,n_{W}, \right\}\) as the news set, where \(V\) and \(W\) are the number of users and news respectively. Each \(u_{v}\) is composed of an impression ID, user ID, news click history \(H = \left\{ n_{1},\ldots,n_{j},\ldots,n_{J} \right\}\), and impressions \(I = \left\{ n_{1},\ldots,n_{k},\ldots,n_{K} \right\}\), where \(J\) and \(K\) are the length of user’s click history and candidate new. The goal of news recommendation is to predict the next click \(n_{J+1}\).
\subsection{Candidate News Encoder}
The candidate news encoder is utilized to learn candidate news representations from news categories, subcategories, titles, and generated titles derived from news abstracts.
\subsubsection{Representation Learning.} Considering the incorporation of multiple fields of news information, in which news categories and subcategories are represented categorical data, and news titles and abstracts contain textual information. We introduce a straightforward template \(X_{prompt}\) to concatenate news categories and subcategories, aligning disparate field data. The formulation of \(X_{prompt}\) is as follows:
\begin{equation}
n_{Cats}^{w} = n_{category}^{w} + \text{``about''~} + n_{Subcategory}^{w}
\end{equation}
Through the aforementioned processing approach, we obtain the news field information denoted as \(n_{Cats}^{w}\), \(n_{Title}^{w}\), and \(n_{Abs}^{w}\), undergo representation learning using the pretrained language model RoBERTa-base\footnote{https://huggingface.co/roberta-base}, resulting in the corresponding representations \(q_{Cats}^{w}\), \(q_{Title}^{w}\), and \(q_{Abs}^{w}\).
\subsubsection{Multi-Field Merging.} Considering the potential variability in the importance of different field information for representing news articles, we aim to allocate attention based on this variability. For instance, in Fig.~\ref{fig1}, the abstract content may be more crucial than the title for conveying factual details. To address this, we utilize additive attention to dynamically learn the weights of different fields across news articles.

Concatenating the representations of multi-field information \(q_{Cats}^{w}\), \(q_{Title}^{w}\), and \(q_{Abs}^{w} \in \mathbb{R}^{b \times d}\) into \(q^{w} = \left\lbrack q_{Cats}^{w},~q_{Title}^{w},~q_{Abs}^{w} \right\rbrack \in \mathbb{R}^{b \times 3 \times d}\),  we calculate the additive attention weights \(\alpha_{i}^{w}\) as follows:
\begin{equation}
\alpha_{i}^{w} = {{e_{w}^{T}\text{~tanh}}\left( {{q^{w}  W}_{p} + b_{p}} \right)},
\end{equation}
\begin{equation}
{\overset{\sim}{\alpha}_{i}^{w}} = \frac{\exp\left( \alpha_{i}^{w} \right)}{\sum\limits_{j = 1}^{n}{\exp\left( \alpha_{j}^{w} \right)}},
\end{equation}
where \(W_{p} \in \mathbb{R}^{d \times h}\), \(e_{w} \in \mathbb{R}^{h \times 1}\) and \(b_{p}\) are learnable parameters. These projections enhance the model's flexibility in attending to information from different news fields. The news representation \({q^{w}}^{'}\) is then computed as a weighted sum of the field representations based on the obtained attention scores \(\alpha_{i}^{w}\):
\begin{equation}
{q^{w}}^{'} = {\sum\limits_{i = 1}^{n}{\overset{\sim}{\alpha}}_{i}^{w}}q_{i}^{w}.
\end{equation}

\subsection{User Encoder}
The user encoder module is used to learn the representations of users from their clicked history. Take user \(u_{v}\) as an example, his clicked history can be donated as \(H^{v} = \left\{ n_{1},\ldots,n_{j},\ldots,n_{J} \right\}\), for each \(n_{j}\) consisting of categories \(n_{Cats}^{j}\), title \(n_{Title}^{j}\), and abstract \(n_{Abs}^{j}\).
    
\subsubsection{Multi-Field Knowledge Extraction.} Considering categories, titles, and abstracts come from distinct fields with unique semantic representation spaces in pretrained language models. And the importance of different fields of information varies. Integrating disparate news field information prematurely may overlook crucial details and introduce noise. For example, a sports enthusiast with a focus on NBA events may heavily weigh news categories as field-specific information. To capture field-level preferences, we conduct multi-field knowledge extraction before merging information.

Specifically, given the clicked news history \(H^{v} = \left\{n_{1},\ldots,n_{j},\ldots,n_{J} \right\}\) of user \(u_{v}\). We divide it into \(H_{Cats}^{v}\), \(H_{Title}^{v}\), and \(H_{Abs}^{v}\). Taking \(H_{Cats}^{v}\) as an example, it is a historical click sequence composed of the categories information for each news \(n_{j}\) in the user's clicked history, donated as \(H_{Cats}^{v} = \left\{n_{Cats}^{1},\ldots,n_{Cats}^{j},\ldots,n_{Cats}^{J} \right\}\). We first obtain its semantic representation \(H_{Cats}^{v}=\left\{q_{Cats}^{1},\ldots,q_{Cats}^{j},\ldots,q_{Cats}^{J} \right\}\) through a pretrained language model. Subsequently, we subject it to field-level multi-head self-attention operations. We employ scaled dot-product attention as the attention function for acquiring the weighting coefficients of the output:
\begin{equation}
Attention\left( {Q,K,V} \right) = \text{softmax}\left( \frac{QK^{T}}{\sqrt{d_{k}}} \right)V,
\end{equation}
where \(Q\), \(K\) and \(V\) represents the query, key and value matrix respectively. The process involves applying \(h\) attention functions concurrently to generate output representations that are subsequently concatenated:
\begin{equation}
{head}_{i} = Attention\left( {QW_{i}^{Q},KW_{i}^{K},VW_{i}^{V}} \right),
\end{equation}
\begin{equation}
MultiHead\left( {Q,K,V} \right) = \left\lbrack {{head}_{1};\ldots;{head}_{h}} \right\rbrack W^{O}.
\end{equation}
where \(W_{i}^{Q}\), \(W_{i}^{K}\) and \(W_{i}^{V}\) are the attention weights of \({head}_{i}\), \(W^{O}\) is the weight of the total concatenation of all attention heads.
\subsubsection{User Preference Extraction.} Through the upper-layer network, we obtained representations \(H_{Cats}^{v}\), \(H_{Title}^{v}\), and \(H_{Abs}^{v}\) containing field-level knowledge for each field. Subsequently, we employ an additive attention network to merge the multi-field information of each news, resulting in a historical sequence \(H^{v} = \left\{ q^{1},\ldots,q^{j},\ldots,q^{J} \right\}\), representing a comprehensive semantic view of each news.

Next, we apply an additive attention network on the historical sequence \(H^{v}\) to model the ultimate user interest representation:
\begin{equation}
\beta_{i}^{j} = {{e_{j}^{T}\text{~tanh}}\left( {{q^{j} \times W}_{q} + b_{q}} \right)},
\end{equation}
\begin{equation}
{\overset{\sim}{\beta}}_{i}^{j} = \frac{\exp\left( \beta_{i}^{j} \right)}{\sum\limits_{n = 1}^{J}{\exp\left( \beta_{n}^{j} \right)}},
\end{equation}
\begin{equation}
q^{v} = {\sum\limits_{i = 1}^{J}{\overset{\sim}{\beta}}_{i}^{j}}q^{i}.
\end{equation}
where \(W_{q} \in \mathbb{R}^{d \times h}\), \(e_{j} \in \mathbb{R}^{h \times 1}\) and \(b_{q}\) are learnable parameters, \({\overset{\sim}{\beta}}_{i}^{j}\) represents the attention weight, and \(q^{v}\) is the historical representation of the \(j_{th}\) user.

\subsection{Click Prediction}
We use the dot-product operation to calculate the probability of user interest in candidate news. Given candidate news representation vector \({q^{w}}^{'}\) and user representation vector \(q^{v}\), the calculation is:
\begin{equation}
\hat{y} = {q^{v}}^{T}{q^{w}}^{'},
\end{equation}
For each impression \(I\), positive samples are the clicked news by the user, and \(K\) news without clicks are randomly sampled as negative samples. And aim to address the imbalance between positive and negative samples, we introduce the focal loss function (\(FL( \cdot )\)) \cite{focalloss}, which includes weighting factors \(\alpha_{i}\) and \(\gamma\) to balance the importance of positive and negative samples.The click prediction loss on a sample is formulated as follows:
\begin{equation}
p_{i} = \frac{\exp\left( {\hat{y}}_{i}^{+} \right)}{{\mathit{\exp}\left( {\hat{y}}_{i}^{+} \right)} + {\sum\limits_{j = 1}^{K}{\exp\left( {\hat{y}}_{i,j}^{-} \right)}}},
\end{equation}
\begin{equation}
L_{Rec} = FL\left( p_{i} \right) = - {{\alpha_{i}\left( {1 - p_{i}} \right)^{\gamma}}}\log\left( p_{i} \right).
\end{equation}
where \({\hat{y}}_{i}^{+}\) and \({\hat{y}}_{i,j}^{-}\) are the predicted click scores for a clicked news and its associated \(j_{th}\) skipped news. \(\alpha_{i}\) and \(\gamma\) are the weighting factors of focal loss function.

\subsection{Cross-Field Contrastive Learning}
To address potential clickbait issue in news titles, we perceive them as a type of noise in obtaining the ultimate representation of news. Inspired by recent advancements in self-supervised learning and contrastive learning techniques \cite{Method_4_cl_paper}, we have devised a cross-field contrastive learning module to mitigate clickbait issue in news titles. Specifically, we treat titles and abstracts as two views of one news article, contrasting their semantic variances to combat clickbait problem.

We formulate a contrastive learning framework to direct self-supervised learning of field-level features to enhance the mutual information between title and abstract information. Technically, for all impressions and clicked news history within a given batch, we conduct cross-field contrastive learning respectively. For a batch \(I_{batch} = \left\{ n_{1}^{1},\ldots,n_{1}^{J},n_{2}^{1}\ldots,n_{2}^{J},\ldots,n_{b}^{1},\ldots,n_{b}^{J} \right\}\), where \(J\) denotes the number of news in an impression and \(b\) represents the batch size. We first compute title and abstract representation vectors \(q_{Title}^{i}\) and \(q_{Abs}^{i}\) for each news item using a feed-forward network. These vectors are then normalized. To achieve title debiasing, we treat different views of the same news as positive samples and different views of distinct news within the same impression as negative samples. The contrastive loss is calculated as follows:
\begin{equation}
L_{CL} = - \frac{1}{I_{batch}}{\sum\limits_{n_{i} \in I_{batch}}{\log\frac{\exp\left( {q_{Title}^{i}}^{T} \cdot {q_{Abs}^{i}}\slash\tau \right)}{\sum\limits_{n_{j} \in I_{batch}}{\exp\left( {q_{Title}^{i}}^{T} \cdot q_{Abs}^{j}\slash\tau \right)}}}}.
\end{equation}
where \(I_{batch}\) is a set of the impression news in the current batch, \({n_{i}}\) is one news in this batch which is composed of \({q_{Title}^{i}}\) and \({q_{Abs}^{i}}\), and \(\tau\) is a hyper-parameter to control the discrimination against negative samples \cite{InfoNCE}.

\subsection{Model Training}
% As mentioned earlier, to alleviate clickbait issue in candidate news, we introduce a contrastive learning loss on top of the original recommendation loss to aid in optimizing our proposed TDNR-C² model. So the overall objective function is formulated as follows:
The final unified loss \(L\) is a weighted summation of these two loss functions, which is formulated as follows:
\begin{equation}
L = L_{Rec} + \lambda_{1}L_{CL}.
\end{equation}
where \(\lambda_{1}\) is the hyper-parameter to control the contribution of the contrastive learning module.

% 第四章：实验
\section{Experiments}
In this section, we compare the proposed TDNR-C² with several competitive methods to showcase the performance of our approach. Furthermore, through ablation experiments, we verify the impact of key components in our proposed TDNR-C² model on recommendation performance. Lastly, we illustrate the effectiveness of TDNR-C² in addressing clickbait issue through a simple example.

\subsection{Experimental Settings}
\subsubsection{Dataset for Experiments.} We validate our model on the real-world dataset MIND\footnote{https://msnews.github.io/.} \cite{MIND} which was proposed by Microsoft in 2021 for news recommendation. The dataset is derived from anonymous user behavior logs on the Microsoft News website. 
To evaluate the performance of our model, we conduct the experiments on the MIND-small dataset and utilize the evaluation metrics AUC, MRR, NDCG@5, and NDCG@10, as specified in the MIND paper, to assess models' effectiveness. We randomly sampled 10\% of training data for validation.
% Each news item consists of news categories, subcategories, titles, abstracts, bodies, and entities.

\subsubsection{Methods for Comparison.} We compared the TDNR-C² model with the following baseline methods: DKN \cite{DKN}, NPA \cite{NPA}, FIM \cite{FIM}, NRMS \cite{NRMS}, LSTUR \cite{LSTUR}, NAML \cite{NAML}, AMM \cite{AMM}, MMRN \cite{MMRN}, MCCM \cite{MCCM}. DKN, FIM and AMM are semantic matching models. NPA, NRMS and NAML are attention based models. LSTUR is LSTM based model. MMRN and MCCM are models based on diverse user modeling techniques.

\subsubsection{Parameter Settings.} We apply negative sampling with a ratio of 4, standardize the title length to 20 for all methods, set the generated title length to 25 for TDNR-C², and for methods incorporating abstract or body content, the length is configured to 50. The maximum length of the user click sequence is set to 25. For the optimizer, we choose to use the Adam optimizer and set the learning rate to 2e-5. In addition, the hyperparameter \(\alpha\) and \(\gamma\) in the focal loss is set to 0.25 and 2, while the fusion ratio \(\lambda_{1}\) and \(\tau\) in contrastive learning is set to 0.1.

\begin{table}
\centering
\caption{The overall performance of different methods on MIND-small. Improvements of AUC over the best baseline (*) are statistically significant with $t$-test ($p < 0.01$).}\label{tab1}
% \caption{The performance of different methods on MIND-small. The best results are marked in \textbf{bold font}, while the second best results are highlighted using \underline{underline}.}\label{tab1}
\renewcommand{\arraystretch}{1}
\setlength{\tabcolsep}{4pt}
\begin{tabular}{l|cccc}
\hline
Method & AUC & MRR & NDCG@5 & NDCG@10 \\ \hline
% DeepFM & 57.94 & 23.95 & 25.22 & 32.21 \\
% Wide\&Deep & 59.94 & 25.35 & 27.43 & 33.27 \\
% DSSM & 62.02 & 26.71 & 28.97 & 34.89 \\
DKN & \(62.85^{\ast}\) & 28.42 & 31.07 & 37.48 \\
NPA & \(64.65^{\ast}\) & 30.02 & 33.11 & 39.41 \\
FIM & \(65.32^{\ast}\) & 30.67 & 33.67 & 40.22 \\
NRMS & \(65.63^{\ast}\) & 30.99 & 34.16 & 40.60 \\
LSTUR & \(65.82^{\ast}\) & 30.76 & 33.92 & 40.11 \\
NAML & \(66.10^{\ast}\) & 31.57 & 34.92 & 41.12 \\
MMRN & \(67.79^{\ast}\) & 32.93 & \underline{36.90} & 42.50 \\
MCCM & \(67.95^{\ast}\)  & 32.76  & 36.62 & 42.66 \\
AMM & \underline{\(67.96^{\ast}\)} & \underline{32.98} & 36.64 & \underline{42.77} \\ \hline
\textbf{TDNR-C²} & \textbf{68.89}\textsuperscript{*} & \textbf{33.57} & \textbf{37.23} & \textbf{43.39} \\ \hline
\end{tabular}
\end{table}

\subsection{Performance Evaluation}
According to the experimental results presented in Table~\ref{tab1}, we conclude that despite models such as DKN, FIM and LSTUR being enhanced with additional news field information, their performance is still comparable to methods that solely rely on news titles (NRMS, NPA). This is due to the lack of field-level knowledge extraction, prematurely integration of multi-field information which leads to the inclusion of noise from specific fields in news representation.

Our proposed TDNR-C² model outperforms all baselines under all evaluation metrics. Comparing to models like NAML and AMM that integrate the body or both body and abstract, the Title Debiasing module in TDNR-C² proves remarkably effective. Even with just a 25-character abstract, it can outperform methods using a 50-character setting and both abstract and body. And comparing with MMRN and MCCM, the former utilizes multiple levels and interests to obtain user representation, while the latter employs multi-channels to acquire user representation, yet cannot prevent the impact of clickbait noise on the representation. This underscores the power of our multi-field knowledge extraction and cross-field contrastive learning in combating clickbait in news titles and enhancing the news recommender's efficiency with long text news field information.

\begin{figure}
\centering
\includegraphics[width=0.8\textwidth]{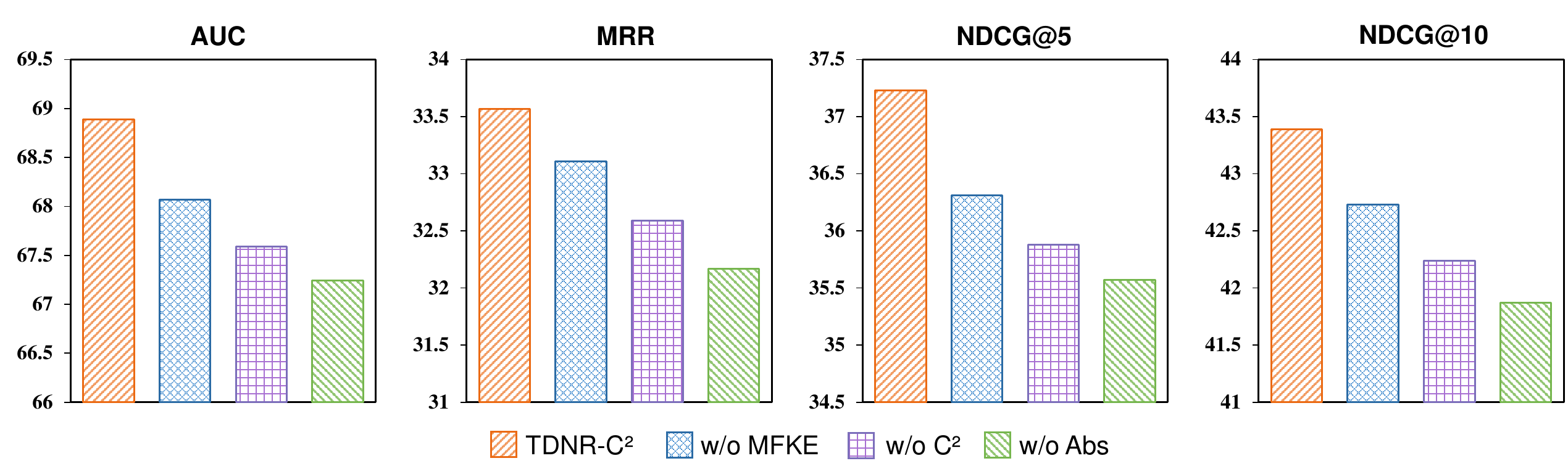}
\caption{Results of ablation experiments.} \label{fig3}
\end{figure}

\subsection{Ablation Study}
To investigate the contribution of each module to the model, we conduct a series of ablation experiments. The results are depicted in Fig.~\ref{fig3}.
\begin{itemize}[label=\textbullet]
\item \textbf{w/o MFKE:} In order to avoid premature integration of knowledge from multiple fields to prevent hidden noise from specific fields affecting news representation, we introduce a multi-field knowledge extraction (MFKE) module to first learn multi-view knowledge from each field before aggregating it for news representation. Comparison between w/o MFKE and TDNR-C² demonstrates the effectiveness of the MFKE module in preventing premature noise integration.
\item \textbf{w/o C²:} To verify that the cross-field contrastive learning module can effectively debias the potential clickbait problem in news title by comparing the title and abstract field information, we propose a variant w/o C², which remove the C² module and set the length of the abstract information from 25 to 50. Comparison between TDNR-C² and w/o C² shows that TDNR-C² outperforms w/o C² by leveraging semantic information in abstract fields to mitigate potential clickbait issues in news titles.
\item \textbf{w/o Abs:} We introduce a variant w/o Abs to assess the impact of abstract field information on performance. This variant generates news representation solely based on news categories and titles. Comparison with w/o C² shows that w/o C² improves recommendation performance by incorporating abstract field information. However, lacking effective title debiasing means, w/o C² does not fully utilize the semantic richness of abstract fields compared to the complete TDNR-C² model.
\end{itemize}

\begin{figure}[!t]
\includegraphics[width=\textwidth]{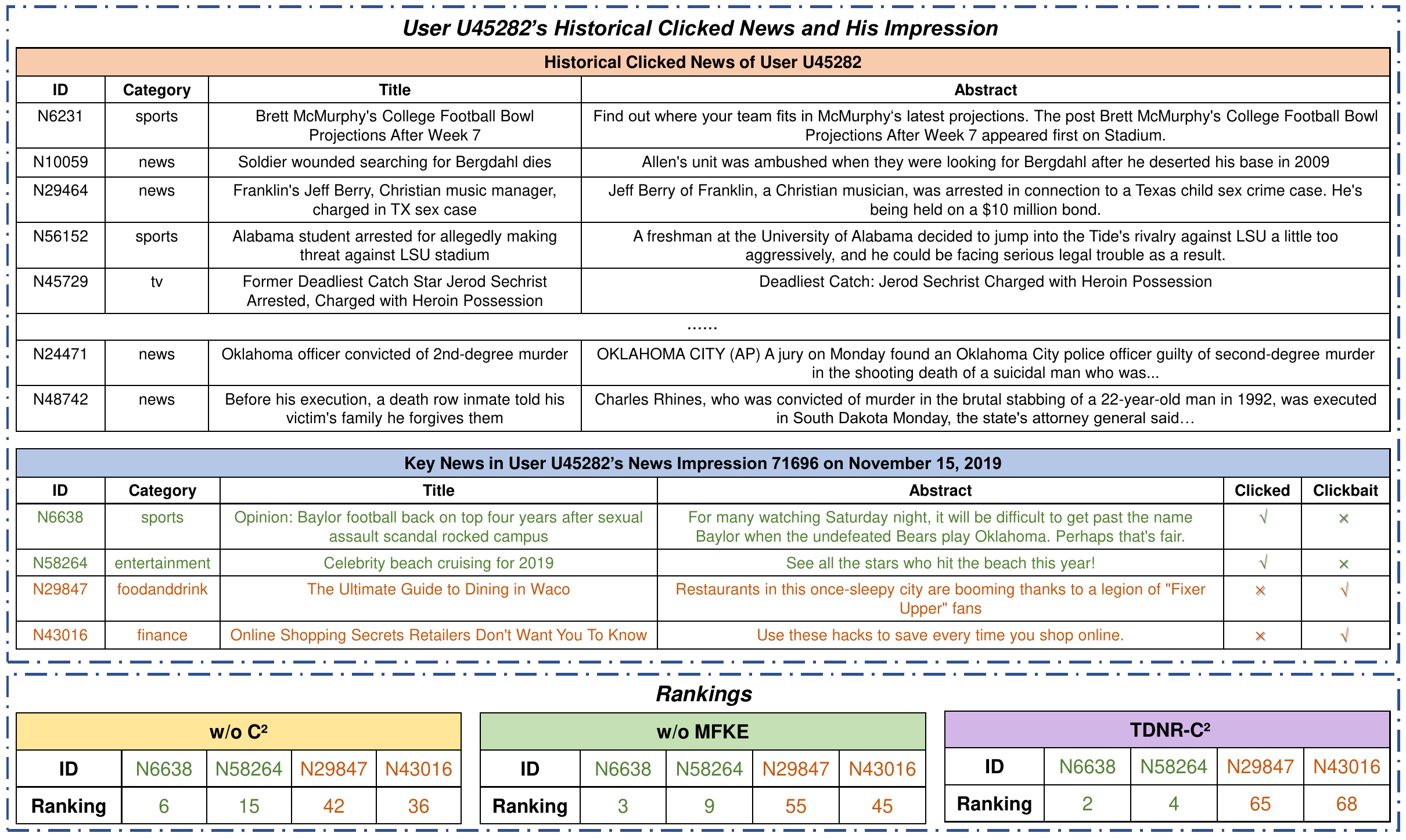}
\caption{Case study of the rankings of truly clicked news and clickbait news.} \label{fig4}
% \caption{Case study of the rankings of truly clicked news and clickbait news by the complete TDNR-C² model and two variants model for impression 71696 of user U45282.} \label{fig4}
\end{figure}
\subsection{Case Study}
To further illustrate the effectiveness of the title debiasing module in mitigating clickbait issue, we conducted a case study. Fig.~\ref{fig4} displays the recommendation rankings of different variant models for news truly clicked by users and news with clickbait issue. This impression consists of 41 news articles clicked by user U45282 and 75 candidate news articles from November 15, 2019. Among these, two were truly clicked, and two had clickbait issues. To visually highlight the clickbait issue in the candidate news, we annotate the title and abstract information. 

The TDNR-C² model outperforms other variants by ranking truly clicked news higher and clickbait news lower. Removing the multi-field knowledge extraction or cross-field contrastive learning modules compromises the system's ability to penalize clickbait articles effectively, impacting performance. This confirms that through multi-field knowledge extraction and cross-field contrastive learning, our model effectively leverages abstract field information to debias news titles, addressing potential clickbait issues.
 % 第六章：结论
\section{Conclusion}
In this paper, we investigate the clickbait issue in news recommendation, noting the value of abstract field data in addressing this challenge. Introducing the innovative TNDR-C² Title Debiasing framework with MFKE and C² modules, we leverage multi-view knowledge and semantic matching between abstract and title fields to combat clickbait issue.
 % MFKE learn the multi-view knowledge of news to prevent noise from propagating between fields. C² is used to match the semantic information in the abstract and title fields to reduce the clickbait problem within news titles.
Experimental results confirm the efficacy of the TDNR-C² model in clickbait mitigation and news recommendation accuracy enhancement. Moving forward, we will delve deeper into exploring the mitigating effects of knowledge from other news fields on the clickbait issue.

\section{Limitations}
Although we have investigated the effectiveness of our model in addressing clickbait through case studies, our analysis is currently limited to the algorithmic level and lacks enhanced interpretability for users. Future research may involve the development of a new component specifically designed to elucidate the decision-making process of the model, particularly in determining whether a news title exhibits clickbait. Such an approach intends to boost users' trust in the recommendation system.

\bibliographystyle{splncs04}
\bibliography{TDNR-C2}

\end{sloppypar}
\end{document}